\documentclass[preprint,10pt, 3p, times]{elsarticle}

\usepackage{adjustbox, booktabs}
\usepackage{lscape}
\usepackage{url, amsmath} 

\usepackage{amssymb}

\journal{Computer Speech \& Language}

\begin{document}
\begin{frontmatter}



\title{Speech Self-Supervised Representations Benchmarking: a Case for Larger Probing Heads}

\author[telecom]{Salah Zaiem}
\ead{salah.zaiem@telecom-paris.fr}

\author[capgemini]{Youcef Kemiche}

\author[samsung]{Titouan Parcollet}

\author[telecom]{Slim Essid}
\author[concordia]{Mirco Ravanelli}

\affiliation[telecom]{organization={LTCI, Télécom Paris},
             addressline={19 Place Marguerite Perey},
             city={Palaiseau},
           postcode={91120},
             country={France}}

\affiliation[capgemini]{organization={Capgemini, HI! Paris Engineering Team},
             city={Paris},
             addressline={11 Rue de Tilsitt},
             postcode={75017},
             country={France}}

\affiliation[samsung]{organization={Samsung AI Center Cambridge},
             addressline={50/60 Station Road},
             city={Cambridge},
             postcode={CB1 2JH},
             country={United Kingdom}}
             
\affiliation[concordia]{
             organization={Concordia University, Mila-Quebec AI Institute},
             city={Montréal},
             addressline={1455 Blvd. De Maisonneuve Ouest},
             postcode={H3G 1M8},
             country={Canada}}



\begin{abstract}
Self-supervised learning (SSL) leverages large datasets of unlabeled speech to reach impressive performance with reduced amounts of annotated data. The high number of proposed approaches fostered the emergence of comprehensive benchmarks that evaluate their performance on a set of downstream tasks exploring various aspects of the speech signal. However, while the number of considered tasks has been growing, most proposals rely upon a single downstream architecture that maps the frozen SSL representations to the task labels. This study examines how benchmarking results are affected by changes in the probing head architecture. Interestingly, we found that altering the downstream architecture structure leads to significant fluctuations in the performance ranking of the evaluated models. Against common practices in speech SSL benchmarking, we evaluate larger-capacity probing heads, showing their impact on performance, inference costs, generalization, and multi-level feature exploitation. \end{abstract}

\end{frontmatter}
\section{Introduction}
Self-supervised learning (SSL) offers a compelling solution for benefiting from abundant unlabeled data to achieve notable performance improvements in various downstream tasks such as speech or speaker recognition. Numerous techniques have been introduced in the literature, such as predictive coding  \cite{baevski2020wav2vec, hubert}, multi-task learning \cite{zaiem2022pretext, ravanelli2020multitask}, and contrastive learning approaches \cite{cola, zaiem2022automatic}. Recently, self-supervised representations have emerged as indispensable tools for speech practitioners who face challenges due to insufficient annotations across an expanding range of tasks \cite{9801640}.

However, experimenting with large SSL models is a costly endeavor both in terms of time and computing. The proliferation of approaches for speech SSL \cite{Mohamed2022} has, therefore, fomented the need for ``universal'' benchmarks evaluating their performance across multiple downstream tasks. These benchmarks should serve as a means to explore different facets of the speech signal, enabling practitioners to make informed decisions tailored to their specific use cases. Benchmarks also allow the research community to have a common field of comparison for the different proposed SSL techniques and identify areas for improvement. Consequently, there has been a growing proliferation of comprehensive benchmarks in recent years \cite{superb, superb_slt, evain2021lebenchmark}. These benchmarks offer standardized frameworks for evaluating the effectiveness of speech SSL models and algorithms. They encompass a wide array of speech applications. Even within a single objective like automatic speech recognition (ASR), they provide various linguistic, acoustic, and prosodic configurations \cite{tsai-etal-2022-superb}.

In prevalent speech SSL benchmarks, the evaluation of self-supervised representations typically involves using downstream decoders that map the frozen representations to the final downstream labels. These downstream probes are generally chosen based on simplicity and limited capacities, such as linear probing for classification tasks or shallow vanilla recurrent neural networks for speech recognition \cite{superb}. However, we hypothesize that this benchmarking approach may harm the development of novel SSL technologies in two significant ways. Firstly, the popularity of the main benchmarks, such as SUPERB \cite{superb}, has established the considered downstream probes as the standard evaluation setting for any new speech SSL model. The metrics used in these benchmarks also contribute to shape the development of new approaches. Consequently, there may be a tendency to discard models that perform poorly with the selected probes, even if they could potentially excel with other downstream architectures. Secondly, the simplicity of the probes contrasts with the increasing complexity of SSL encoders. Testing with low-capacity probes can lead to an unnecessary transfer of complexity from the probing head, which is intended to be task-specific, to the encoder, which is expected to be more general. This transfer can result in unnecessarily large self-supervised models, leading ultimately to compute-costly inferences \cite{zaiem2023fine}. For example, in computer vision, Dubois et al. \cite{dubois2022improving} demonstrated that changing the probe family from linear to multi-layer perceptrons (MLP) leads to different optimal hyperparameter values of SSL models and enables smaller SSL representations.

One potential solution to address these limitations is to explore headless evaluation alternatives that are not tied to specific downstream probes. While a few intrinsic quality assessment metrics for speech embeddings have been proposed \cite{schatz}, their correlation with downstream performances is still uncertain \cite{algayres2020evaluating}. In image classification, Garrido et al. \cite{garrido2022rankme} demonstrated a strong correlation between the rank of vision SSL representations and final downstream performance, though  the latter performance is obtained using linear probes exclusively. Recognizing these challenges, SUPERB \cite{superb} offers two tracks where researchers can choose their own downstream probes, with or without capacity constraints on the probing architectures. Regrettably, these two tracks have yet to receive any submissions.

This paper builds on previously published findings \cite{zaiem2023speech} which diagnosed the dependence of benchmarks on the choice of probing heads. Given that our initial results showed that different probing heads lead to different rankings, we argue that it is important to re-question the current practice followed by prominent benchmarks, where a particular probe is fixed for each task, without a clear justification.  In this sense, we extend our previous study with a more thorough assessment of the benefits of performing the benchmarks with more-capacitated probing heads. Precisely, four desired characteristics are assessed: full pipeline performance, inference efficiency, generalization ability, and the exploitation of multi-level encoder features. On all these points, our study shows an advantage for higher-capacity probing heads. These ideas and results aim to reshape the way the SSL models are benchmarked, and indirectly, ultimately influence their design towards better rankings in these benchmarks. Hence, the contributions of this work are fourfold:   

\begin{enumerate}   
    \item We benchmark a set of published state-of-the-art SSL models on various speech tasks, varying the downstream decoders, showing that, except for ASR on Librispeech, the rankings and relative performance are highly impacted by a change in the set of downstream probes (Section \ref{sec:setting}).
    \item We provide an extensive study on the impact of selecting higher-capacity decoders on performance, generalization abilities, inference efficiency, and feature-level selection and exploitation (Sections \ref{sec:results} and \ref{sec:capacity}).
    \item We show that current ``headless" evaluation methods, based on triplet mining, are not an acceptable alternative for self-supervision benchmarking as their results correlate poorly with performance obtained with the best downstream heads (Section \ref{sec:headless}).
    \item We release the code base developed within the SpeechBrain library \cite{speechbrain} for replication and to encourage further investigations and comparisons between models.\footnote{\url{github.com/speechbrain/benchmarks/tree/main/benchmarks/MP3S}} The clean and easy-to use code is released within the ``Benchmarks" SpeechBrain sub-library. We call it ``MP3S" standing for ``Multi-Probe Speech Self-Supervision".
\end{enumerate}

\section{Benchmarking SSL Models: Definition and Protocol}
\label{sec:setting}
This section formally describes the limitations faced by current speech SSL benchmarks and also details the experimental protocol devised to bring this issue to light.

\subsection{Problem definition}
Formally, an SSL pipeline consists of two systems: a pre-trained encoder $\phi$ and a downstream probe $f$. $\phi$ is learned through solving a pretext task on unlabeled speech datasets (e.g., Libri-light \cite{Kahn_2020} and LibriSpeech \cite{libri} have been popular choices in the literature), while $f$ is learned for a considered downstream task with its corresponding annotated training dataset. In this framework, the SUPERB benchmark has chosen  a probing family $\mathfrak{F}_T$ (\textit{i.e.} a downstream architecture with its hyperparameters, such as an MLP with given number of layers and hidden sizes) for every considered downstream task $T$ and, for every considered SSL encoder $\phi$, it shows a task error rate equal to: 
\begin{equation}
\label{eq:superb}
\min\limits_{f \in \mathfrak{F}_T} E_t(f\circ \phi) ;
\end{equation}
with $E_t(f\circ \phi)$ being the test-set error rate of the SSL pipeline. 

However, ideally, as proposed in the \textit{``unconstrained"} track of SUPERB \cite{superb}, the shown performance should be:
\begin{equation}
    \min_{\mathfrak{F} \in \mathfrak{P}} \min_{f \in \mathfrak{F}} E_t(f\circ \phi);
\end{equation}
with $\mathfrak{P}$ the set of all probing families. More interestingly, in the \textit{``constrained"} scenario, if we denote by $\mathfrak{C}$ the set of probes that respect a chosen capacity constraint, then the performance of an encoder $\phi$ could be expressed as follows: 
\begin{equation}
\label{eq:constrained}
    \min_{\mathfrak{F} \in \mathfrak{P} } \min_{f \in \mathfrak{F} \cap \mathfrak{C}} E_t(f\circ \phi).
\end{equation}

 Unfortunately, this quantity cannot be computed, as it would require training a model with every known downstream architecture that respects capacity constraints, for each considered encoder and task. 
 
 In this study, we aim to investigate whether benchmarking based on the value obtained in Equation \eqref{eq:superb} provides a robust ranking that remains consistent across different probing families. To achieve this, we examine different probing families for each downstream task and analyze whether the rankings and relative differences obtained in the initial experiments remain consistent in the subsequent experiments.

\subsection{Self-supervised pretrained models} 

For our study, we focused on a subset of state-of-the-art models from the SUPERB benchmark due to their wide adoption within the community. We selected nine SSL models that extract representations directly from the waveform: Wav2vec 2.0 \cite{baevski2020wav2vec}, HuBERT \cite{hubert}, WavLM\footnote{We used the Base+ version of WavLM, trained on $94k$ hours of speech data} \cite{Chen2021}, and Data2Vec \cite{baevski2022data2vec} in both their Base and Large versions. We also included DistilHuBERT \cite{Chang2021}, which is a distilled version of Hubert Base with four times fewer transformer layers. These models share the same frame rate, generating representations of dimension $D$ every 20 ms of audio signal. $D=1,024$ for the  ``Large" versions and $D=768$ for ``Base" ones and DistilHuBERT.

These models share similar Transformer-based architectures, but their pretraining pretext tasks vary. Wav2vec2.0 is trained using contrastive predictive coding (CPC), aiming to maximize mutual information between contextual features and predicted future samples. HuBERT and WavLM learn to map unlabeled audio to sequences of pseudo-labels generated through clustering previously generated representations. WavLM introduces training distortions to HuBERT enabling noise-invariant representations. Data2Vec, inspired by teacher-student approaches, employs a masked input view to predict latent representations of the unmasked input data, utilizing a self-distillation setup. We obtained all the pre-trained checkpoints from their respective HuggingFace (HF) official cards \cite{wolf2020transformers}, except for Wav2vec2.0 Large, for which we used the Fairseq \cite{ott2019fairseq} checkpoint since the HF version underperformed compared to the results reported in SUPERB.

\subsection{Downstream Tasks and Datasets}
Speech SSL benchmarks attempt to assess universal speech representations by offering a diverse array of tasks that examine various facets of the speech signal. In line with this approach, we introduce seven tasks that cover phonetic, speaker-identity, emotional, and semantic dimensions. \\

\noindent \textbf{Speech Recognition Tasks.} Four speech recognition tasks are considered. For the first one, LibriSpeech \cite{libri} \textit{train-clean-100}/\textit{dev-clean} subsets are used for training and validation while \textit{test-clean} and \textit{test-other} are kept for testing. The Buckeye dataset \cite{buckeye} is considered as a second ASR task, allowing for testing the ability of the models with fewer labeled data and in a more complex spontaneous setting of English speech. The training, validation, and test splits used in our Buckeye experiments are available in the companion repository with the training set containing approximately $9.5$ hours of audio and the test set $1.5$ hour. 
For these two English ASR tasks, we present two sets of results based on the use or not of a language model (LM) during the decoding process. In the experiments labeled ``Without LM," we employ greedy decoding. Conversely, the ``With LM" experiments utilize the official LibriSpeech 4-gram language model combined with shallow fusion to the acoustic model. Since low-resource languages are one of the main applications of SSL methods, two low-resource language tasks, extracted from the CommonVoice $11.0$ \cite{ardila2020common} release, are considered: Welsh (\textit{Cymraeg}) and Basque (\textit{Euskera}). To ease reproducibility, we use the splits provided in the CommonVoice release: the Basque train set is 15.8-hour long, with 56 different speakers, while test and dev splits are 10.5 and 9.8-hour long. For Welsh, train, dev and test, splits are respectively, 11, 7.9 and 8 hour-long with 32 different speakers in the training set. The Word Error Rate (WER) serves as the error metric for all ASR tasks. In all experiments, the probe is trained using the Connectionist Temporal Classification (CTC) loss at the character level. \\

\noindent \textbf{Automatic Speaker Verification (ASV).} The ASV task consists of a binary classification procedure aimed at determining whether speakers in a pair of utterances are the same. Similar to the SUPERB benchmark, we utilize the VoxCeleb1 train and test splits for this task \cite{Nagrani_2017}. It is worthwhile to note that the testing set may include speakers who were not present in the training set. The evaluation metric employed for ASV is the Equal Error Rate (EER). \\

\noindent \textbf{Emotion Recognition (ER).} For ER, we utilize the IEMOCAP dataset \cite{busso2008iemocap}, which comprises $10,039$ utterances from $10$ distinct speakers. The objective of this task is to predict the emotional class of a speech  utterance from four possible candidates: \textit{neutral}, \textit{happy}, \textit{sad}, and \textit{angry}. The reported performance represents the mean of 10 runs conducted through cross-validation on 10 folds, where each fold leaves out the data of one speaker for testing purposes. \\

\noindent \textbf{Intent Classification (IC)}. While the SUPERB benchmark evaluates the semantic content of SSL representations using the Speech Commands (SC) \cite{warden2018speech}, we employ the more challenging SLURP dataset \cite{bastianelli2020slurp} for Intent Classification, as error rates with SC are extremely low. The SLURP collection consists of approximately $72,000$ audio recordings that capture user interactions with a home assistant in single-turn scenarios. The IC task involves classifying each utterance into one of the 18 predefined scenarios, such as ``calendar", ``email", and ``alarm". Classification accuracy serves as the metric for both emotion recognition and intent classification tasks.

\subsection{Downstream Probes}\label{probes}
This section offers a high-level description of the downstream probes employed in the study. For comprehensive replication of the experiments, detailed information regarding hyperparameters and architectural specifications can be found in the code repository.\\

\noindent \textbf{Global settings.} During the downstream training, the weights of the SSL encoder are kept frozen, learning solely the weights of the downstream decoder. Similarly to SUPERB, we observed that the last-layer representation may not always be optimal. Consequently, we, first, store the representations from all hidden layers of the pre-trained model. These hidden states are then weighted and summed to create the representation forwarded to the decoder. The weights are trained during the downstream process. In order to ensure the validity of our experimental setting, we first reproduced the downstream architectures used in SUPERB during the initial set of experiments. Then, we modified the probes by introducing simpler or more complex alternatives inspired by the relevant literature for each task. 

\noindent \textbf{Speech recognition tasks.} In the initial set of experiments, aimed at replicating the SUPERB conditions, a vanilla 2-layer Bidirectional LSTM (BiLSTM) with $1,024$ units is utilized. This BiLSTM is followed by a linear layer that maps the latent representations to characters. For the second set of downstream architectures, we employ an encoder-decoder Conformer architecture \cite{gulati2020conformer} for the LibriSpeech task. The downstream architecture consists of $12$ encoder layers, $4$ decoder layers, and $4$ attention heads. For the Buckeye task, we employ the convolutional-based ContextNet architecture \cite{han2020contextnet} with unit strides to maintain the frame rate of the SSL models. In the case of Welsh and Basque from CommonVoice, a two-layer dense neural network is employed to map each frame representation to the probabilities of the corresponding characters. Additionally, experiments using ContextNet with LibriSpeech are also conducted. The performance of ContextNet and Conformer architectures, which are close to the state-of-the-art on LibriSpeech, motivated their selection as downstream probes. Different probes are selected for ASR tasks to show that eventual variations in performance are not linked to a unique couple of probes.\\

\noindent \textbf{Automatic speaker verification.}  In the first experiment, we use the X-vector architecture \cite{snyder2018x} with the AM-Softmax loss \cite{wang2018additive} for training speaker embeddings. Verification is performed using cosine similarity backend. In the second experiment, we employ the ECAPA-TDNN neural network \cite{desplanques2020ecapa}, which integrates time-delay neural networks and attention mechanisms to capture temporal dependencies and achieve state-of-the-art results in speaker verification \cite{desplanques2020ecapa}. \\
\begin{table}[]
\centering
\scalebox{0.95}{
\begin{tabular}{lcc} \toprule
\textbf{Task/Probing Head}             & \textbf{First Set}             & \textbf{Second Set}      \\ \midrule
LibriSpeech ASR                & BiLSTM                & Conformer \cite{gulati2020conformer}       \\
Buckeye ASR                    & BiLSTM                & ContextNet \cite{han2020contextnet}     \\
CommonVoice Low-Resource  ASR  & BiLSTM                & Linear          \\
Automatic Speaker Verification & X-Vectors  \cite{snyder2018x}           & ECAPA-TDNN \cite{desplanques2020ecapa}     \\
Emotion Recognition            & Time-Pooling + Linear & ECAPA-TDNN  \cite{desplanques2020ecapa}      \\
Intent Classification          & Time-Pooling + Linear & BiLSTM + Linear \cite{lugosch2019speech} \\ \bottomrule
\end{tabular}}

\caption{ Probes selected for the downstream trainings. More details can be found in the companion repository. }
\label{tab:probes}
\end{table}
\noindent\textbf{Classification tasks.} Similar to SUPERB, in the initial set of experiments, we employ linear probing for the classification tasks, namely intent classification and emotion recognition. The representations are first averaged along the time axis and then passed through a linear classification layer. For the second downstream architecture, inspired by state-of-the-art approaches  \cite{wang2021fine}, we opt for ECAPA-TDNN for emotion recognition. As for intent classification, we follow published work \cite{lugosch2019speech} and utilize two layers of BiLSTM with a hidden size of $1,024$, followed by a linear classifier. This approach allows for considering the order of frame representations, in contrast to using time-pooled features. While the cited works (\cite{wang2021fine, lugosch2019speech, desplanques2020ecapa}) employ these architectures on-top of handcrafted features (generally log-mel spectrograms), we show in the following that they are still relevant when fed with self-supervised representations.  Table \ref{tab:probes} provides a summary of the probing heads selected for our experiments.


\begin{table*}[]

\centering
\scalebox{0.68}{
\begin{tabular}{lcccccccccccc} \toprule
\textbf{Models /Tasks} & \textbf{SSL Params.} & \multicolumn{4}{c}{\textbf{LibriSpeech train-100 ASR}} & \multicolumn{2}{c}{\textbf{Buckeye ASR}} & \textbf{Welsh} & \textbf{Basque} & \textbf{ASV} & \textbf{ER} & \textbf{IC} \\ \midrule
\multicolumn{2}{l}{Evaluation Metrics}        & \multicolumn{4}{c}{WER $\downarrow$} &      \multicolumn{2}{c}{WER $\downarrow$}                   & WER $\downarrow$           & WER $\downarrow$            & EER $\downarrow$     & Acc. $\uparrow$ & Acc. $\uparrow$ \\ \midrule

\multicolumn{2}{l}{First downstream architectures}        & \multicolumn{4}{c}{LSTM} &      \multicolumn{2}{c}{LSTM}                   & LSTM           & LSTM            & Xvectors     & Pool + Lin. & Pool + Lin. \\ \midrule

 &  & \textbf{Clean} & \textbf{Other} & \textbf{Clean LM} & \textbf{Other LM} & \textbf{w/o LM} & \textbf{with LM} & \textbf{Welsh} & \textbf{Basque} & \textbf{ASV} & \textbf{ER} & \textbf{IC} \\ \midrule
DistilHuBERT           & 23.5M               & 13.99              & 34.91               & 9.96                  & 28.26                 & 35.59            & 28.29               & 53.20          & 46.78           & 9.1        & 65&46.6                 \\
Wav2vec 2.0 Base       & 95M                 & 6.23               & 14.93               & 4.86                  & 11.97                 & 24.87            & 19.48               & 54.45          & 51.21           & 5.29         & 66.4   & 59.0           \\
Wav2vec 2.0 Large      & 317.4M              & 3.72               & 9.25                & 3.13                  & 7.48                 & \textbf{20.72}            & 16.11               & 45.42          & \textbf{37.98}           & 5.69         & 69.3 &66             \\
HuBERT Base            & 94.7M               & 6.24               & 15.03               & 5.03                 & 12.31                 & 45.53            & 26.51               & 52.92          & 46.91           & 4.50         &67.5 & 53.8                \\
HuBERT Large           & 316.6M              & 3.57               & 8.12                & 2.90                  & 6.59                  & 51.30            & 33.10               & 51.21          & 46.15           & 5.20         & 71.3 &69.9              \\
WavLM Base+            & 94.7M               & 5.96               & 14.33               & 4.84                  & 11.72                 & 42.21            & 24.41               & 51.31          & 46.40           & 3.74         & 67.1 & 57.9               \\
WavLM Large            & 316.6M              & 3.48               & 7.37                & 2.87                  & 5.96                  & 27.31            & \textbf{14.27}               & 48.92          & 41.89           & \textbf{2.98}    & \textbf{75.3}      & \textbf{78.8}               \\
Data2vec Base          & 93.8M               & 5.30               & 13.79               & 4.03                 & 10.97                 & 37.26            & 30.50               & 54.00          & 46.37           & 5.43        & 63.0    & 56.9             \\
Data2vec Large         & 314.3M              & \textbf{3.10}            & \textbf{6.50}            & \textbf{2.58}                & \textbf{5.38}                  & 22.63            & 18.63               & \textbf{44.32}          & 38.23           & 4.89      & 64.1   & 69.8                \\ \midrule
\multicolumn{3}{l}{\textbf{Probe size and inference metrics}} \\ \midrule
\multicolumn{2}{l}{Downstream Parameters Base}             & \multicolumn{4}{c}{39.9M}             & \multicolumn{2}{c}{39.9M}                & 40.3M          & 40.3M           & 7.0M         & 13.8k       & 3.1k        \\
\multicolumn{2}{l}{Downstream Parameters Large}               & \multicolumn{4}{c}{42M}                     & \multicolumn{2}{c}{42M}                  & 42.4M          & 42.4M           & 7.7M         & 18.4k       & 4.1k        \\ \midrule  \midrule

\multicolumn{2}{l}{Second downstream architectures}                     & \multicolumn{4}{c}{Conformer} &      \multicolumn{2}{c}{ContextNet}                   & Lin.           & Lin            &  ECAPA      & ECAPA & LSTM + Lin. \\ \midrule
 &  & \textbf{Clean} & \textbf{Other} & \textbf{Clean LM} & \textbf{Other LM} & \textbf{w/o LM} & \textbf{with LM} & \textbf{Welsh} & \textbf{Basque} & \textbf{ASV} & \textbf{ER} & \textbf{IC} \\ \midrule
DistilHuBERT                & 23.5M       & 14.97                              & 36.51                              & 11.54                 & 31.41                 & 58.56                         & 43.61               & 80.78               & 77.04                & 2.85          & 72.4       & 74.9        \\
Wav2vec 2.0 Base            & 95M         & 6.91                               & 15.39                              & 5.09                 & 12.29                 & 30.04                         & 23.04               & 74.31               & 71.76                & 2.82          & 73.2       & 77.7        \\
Wav2vec 2.0 Large           & 317.4M      & 4.32                              & 9.25                              & 3.58                  & 7.03                 & 23.92 & 18.68               & 75.45               & 78.48                & 3.17          & 68.4       & 79.0        \\
HuBERT Base                 & 94.7M       & 6.88                               & 15.68                              & 5.23                  & 12.63                 & 30.44                         & 23.11               & 77.39               & 73.40                & 2.40          & \textbf{78.2}       & 79.4        \\
HuBERT Large                & 316.6M      & 3.96                               & 8.60                               & \textbf{3.10}                  & 6.88                  & 39.39                         & 31.57               & 71.58               & 60.24                & 3.84          & 71.5       & 80.1        \\
WavLM Base+                 & 94.7M       & 6.55                               & 14.93                              & 4.98                  & 11.80                 & 27.73                         & 21.69               & 75.87               & 69.43                & \textbf{1.76}          & 72.6       & 81.2        \\
WavLM Large                 & 316.6M      & 4.08                              & 8.10                               & 3.13                  & \textbf{6.31}                  & \textbf{15.61}                         & \textbf{12.1}                & \textbf{68.73}               & \textbf{56.32}                & 1.77          & 77.4       & \textbf{85.8}        \\
Data2vec Base               & 93.8M       & 5.85                               & 14.32                              & 4.53                  & 12.52                 & 40.53                         & 33.45               & 77.49               & 75.26                & 3.75          & 72.0       & 73.4        \\
Data2vec Large              & 314.3M      & \textbf{3.43}                              & \textbf{6.82}                             & 3.27                  & 6.58                  & 25.26                         & 21.5                & 69.09               & 63.31                & 2.67          & 71.3       & 79.9        \\\midrule
\multicolumn{3}{l}{\textbf{Probe size and inference metrics}} \\ \midrule
\multicolumn{2}{l}{Downstream Parameters Base}                      &\multicolumn{4}{c}{11.2M}                     & \multicolumn{2}{c}{32.4M}                            & 1.9M                & 1.9M                 & 9.2M          & 7.3M       & 42M         \\
\multicolumn{2}{l}{Downstream Parameters Large}                       & \multicolumn{4}{c}{11.2M}                     & \multicolumn{2}{c}{32.5M}                        & 2.3M                & 2.3M                 & 9.8M          & 7.9M       & 44.1M       \\ \bottomrule  
\end{tabular}}
\caption{SSL benchmarking results for all tasks and downstream architectures.  The number of parameters of the SSL encoder and the probes is shown in the ``Params" rows and columns. Upper part corresponds to the results obtained using the first set of probing heads while the bottom part shows these obtained with the second set. Probing heads are compiled in Table \ref{tab:probes}. }

\label{tab:DS}
\end{table*}
\section{Benchmarking Results and Discussion}
\label{sec:results}

Table \ref{tab:DS} presents the comprehensive benchmarking results for the different SSL models. The upper and lower sections of the table display the performance achieved by the first and second sets of downstream architectures, respectively. Additionally, the number of neural parameters is reported for both the SSL encoder and downstream decoders. For the latter, only two values are provided per task (\textit{i.e.},``Base" or ``Large") as this number only depends on the dimension of the encoder output representations ($D = 1024$ for ``Large" and $D=768$ for ``Base"). In the initial set of experiments, we replicated the SUPERB benchmark conditions for two tasks: LibriSpeech and VoxCeleb1. Notably, our results exhibited a Pearson correlation of 0.99 and 0.97, respectively, with the corresponding results on the SUPERB leaderboard. This high correlation validates our successful replication of the benchmark settings.

To study the impact of a decoder change on the final performances, we compute, for every task, the Pearson and Spearman correlations between the performance metrics obtained with the first downstream architectures and those obtained with the second ones, and collect them in Table \ref{tab:correlations}. The Pearson correlation evaluates the linear relationship between the two sets of metrics, while the Spearman one assesses the strength and direction of their monotonic relationship. Correlation metrics close to $1$ imply proportional performances and similar rankings between the SSL models used with different probes, making the benchmark robust to the considered downstream change. Correlation metrics close to zero indicate no correlation between the results of the two sets of experiments.

All the models tested demonstrate competitive performances on every downstream task and with every related decoding architecture. With the notable exception of LibriSpeech, all the downstream tasks error metrics vary substantially with changing probes. The mean performance of the SSL candidates with the first and second downstream decoders is presented in the last three columns of Table \ref{tab:correlations}. Notably, we observe a significant sensitivity to the choice of decoder as replacing the SUPERB decoder results in relative improvements of up to $46.5\%$ for ASV and $27.3\%$ for IC. This demonstrates the substantial impact that the decoder selection has on the performance of the SSL models.  Furthermore, the Spearman and Pearson correlation values computed between the performances with the first and second set of downstream probes are low, despite being positive.  This suggests significant variations in relative performances and rankings when comparing the results obtained with the two different downstream decoders. For instance, the Spearman correlation coefficients for ER and IC are only 0.34 and 0.66, respectively. It is noteworthy that while the assessment of LibriSpeech performance appears to be robust to decoder changes, this does not hold true for other ASR tasks. In the case of the spontaneous English Buckeye corpus, there is a Spearman correlation of 0.56 and a Pearson correlation of 0.42, while the Basque task exhibits correlations, Pearson and Spearman, of only 0.19 and 0.15. The Buckeye ASR scenario is particular as changing the decoder from BiLSTM to ContextNet leads to improved results for some models and detrimental effects for others. Specifically, the best-performing model, WavLM Large using the second decoder, ranks only fourth when evaluated with the SUPERB settings.

\begin{table*}[]

\centering
\scalebox{0.85}{\begin{tabular}{lccccccc} \toprule
\textbf{Task}   & \textbf{Pearson} & \textbf{Spearman} & \textbf{Mean  DS1} & \textbf{Mean DS2} & \textbf{Diff (\%)} & \textbf{FBANKS DS1} & \textbf{FBANKS DS2} \\ \midrule
LibriSpeech 1-2 & 0.99    & 0.97     & 5.8             & 6.48          & -11.7 & 22.56 &  8.91    \\
Librispeech 1-3 & 0.99    & 0.98     & 5.8             & 7.03          & -21.2 & 22.56 & 43.12  \\ 
Buckeye ASR         & 0.42    & 0.56     & 34.16           & 32.39         & 5.2   & 54.17 & 78.90    \\
Welsh           & 0.59    & 0.62     & 50.64           & 74.52         & -47.2 & 99.62  & $>$ 100  \\
Basque          & 0.19    & 0.15     & 44.66           & 69.47         & -55.6 & $>$ 100 & $>$ 100    \\
ASV        & 0.47    & 0.75    & 5.2             & 2.78          & 46.5  & 9.28 & 3.41    \\
ER         & 0.22    & 0.34     & 67.66           & 73            & 7.9  & 48.51 & 65.7     \\
IC           & 0.75    & 0.66     & 62.1            & 79.04         & 27.3 & 12.6 & 42.3     \\
\bottomrule  
\end{tabular}}
\caption{Correlations (Pearson and Spearman) between the performances achieved with the first and second downstream probes are given for each task. The number in the column name indicates whether the results correspond to the first or second set of probing heads, and ``DS" stands for ``Downstream". ``Mean " columns show the mean performance across all the considered SSL encoders. The ``Diff" column presents the relative  difference in mean performance between the two architectures.  The ``FBANKS "  columns show the performance on every task with Mel spectrograms as input representations. The difference between ``Mean DS" and ``FBANKS DS" outlines the performance gain in \% from using SSL representations instead of handcrafted ones.}

\label{tab:correlations}

\end{table*}
However, we noticed a contrasting pattern in the rankings and performance of the considered SSL encoders on the ASR task using LibriSpeech \textit{train-clean-100}, as shown in Table \ref{tab:correlations}. Unlike the other downstream tasks, the rankings and performance only exhibit minor variations when the downstream decoder is changed. To validate this observation, we conducted additional experiments using a third downstream decoder, ContextNet, specifically for this task. The results of this supplementary experiment are presented in Table \ref{tab:DS3}, and the correlation values between performances with the first probe and the ContextNet are shown in the second row of Table \ref{tab:correlations}. Similarly, no significant differences were observed in the ranking of the SSL candidates. For instance, in all three setups without LM decoding, DistilHuBERT consistently exhibits the lowest performance among the candidates. Furthermore, ``Large" versions of the considered candidates consistently outperform their ``Base" counterparts on this task, independently of the used probing head. Table \ref{tab:correlations} further confirms these findings, revealing high Spearman and Pearson correlations exceeding $0.97$ for LibriSpeech, while the highest correlation value observed for other tasks is only $0.75$. This discrepancy indicates that the SSL encoders might be biased towards the LibriSpeech ASR task, which is not unexpected given its prominent role as a benchmark dataset and its consistent inclusion in the pretraining process datasets. These results lead us to the conclusion that current SSL benchmarking is highly dependent on the choice of the downstream probes, with the notable exception of LibriSpeech ASR.

\begin{table}[]
\centering
\scalebox{0.85}{\begin{tabular}{lccccc}

\toprule
\textbf{Tasks \textbackslash Models} & \textbf{SSL  Params} & \textbf{Clean} & \textbf{Other} & \textbf{Clean LM} & \textbf{Other LM} \\ \midrule

DistilHuBERT                         & 23.5M                & 20.52                & 43.27                & 10.44                   & 29.17                   \\
Wav2vec 2.0 Base                     & 95M                  & 7.24                 & 15.66                & 4.73                    & 11.21                   \\
Wav2vec 2.0 Large                    & 317.4M               & 4.35                 & 8.68                 & 03.03                   & 6.86                    \\
HuBERT Base                          & 94.7M                & 7.31                 & 16.00                & 4.60                    & 11.11                   \\
HuBERT Large                         & 316.6M               & 4.04                & 8.63                 & 2.98                    & 6.45                    \\
WavLM Base+                          & 94.7M                & 6.73                 & 15.33                & 4.52                    & 10.84                   \\
WavLM Large                          & 316.6M               & 4.09                & 8.43                 & 2.94                    & 6.15                    \\
Data2vec Base                        & 93.8M                & 5.46                 & 13.34                & 3.76                    & 10.04                   \\
Data2vec Large                       & 314.3M               & \textbf{3.50}                 & \textbf{6.94}                 & \textbf{2.56}                    & \textbf{5.36}                    \\ \midrule
\multicolumn{3}{l}{\textbf{Probe size and inference metrics}} \\ \midrule

\multicolumn{2}{l}{Downstream Parameters Base}                 & \multicolumn{4}{c}{32.4M}         \\
\multicolumn{2}{l}{Downstream Parameters Large}                               & \multicolumn{4}{c}{32.5M}       \\ \bottomrule           
\end{tabular}}

\caption{Word Error Rate (WER \%) results of LibriSpeech experiments on the two considered test splits with Contextnet as a third downstream probe. ``DS" stands for Downstream.}

\label{tab:DS3}
\end{table}

\section{On limited-capacity probing heads} \label{sec:capacity}
The first section has shown that the rankings and relative performances of the benchmarked self-supervised systems are heavily impacted by a change in the downstream probing heads. The question that naturally arises is whether the common choice of probing heads is justified enough to discourage evaluating with other alternatives. The proposed downstream probes in the prominent SUPERB benchmark were selected based mainly on a simplicity criterion. Choosing simple probing heads is generally justified by the fact that it allows for evaluating only the quality of the pre-trained representations and not the downstream probes learning abilities. In this section, we will show that choosing limited-capacity decoders is not optimal. To prove it, and based on the previous experiments and further ones, we will show that larger probing heads: 1) lead to better performance; 2) reduce the error rate gaps between large and smaller SSL encoders, potentially leading to lower inference times; 3) enable the exploitation of multi-level features within the encoders; and 4) do not harm the generalization abilities of the full pipeline.

This subsection elaborates two conclusions from the presented results and further computations of inference metrics. First, on most tasks, larger capacity decoders improve significantly the performance, allowing an optimal use of the pretrained representations. Second, larger-capacity probes enable smaller SSL encoders to bridge the performance gap with larger ones, eventually leading to faster inferences.

Concerning performance, Table \ref{tab:correlations} shows that except for the Buckeye ASR task, the mean performance is better with the probes with larger capacities, mainly for Speaker Verification and Intent Classification with respectively $46.5\%$ and $27.3\%$ relative performance improvements (for ASR tasks, the first probe, two layers of BiLSTM, is the largest probe in terms of number of parameters as shown in Table \ref{tab:DS}).  Decoders with more capacity seem naturally able to better exploit the benchmarked representations. For instance, time-pooling the frame-level representations before emotion or intent classification prevents the model from learning to use local or time-ordered signal clues, while it is possible with ECAPA-TDNN or a layer of BiLSTM in the probing head. To know whether the performance increase is imputable to the representations or the probes,  we compute the performance of the downstream probes using Mel-scaled spectrograms as the input representation. The spectrograms' extraction is done similarly to the one provided as baseline in the SUPERB benchmark \cite{superb}. The results are shown in the last two columns of Table \ref{tab:correlations}. We can see, first, that the mean performance is significantly better using learned representations than hand-crafted Mel spectrograms, especially for ASR where the final WER is over $100$ in three cases. For intent classification, the accuracy using SSL representations, is in average $5x$ better with the first probe and twice higher with the second probe.  Moreover, apart for VoxCeleb, where two models perform worse than spectrograms with the second probe, all the representations benchmarked lead to better performances with all probes on all considered tasks. This shows that the lower error rates reached using larger decoders still depend on the quality of the input representations and that the levels of performance reached allow for an informed ranking of those.

Additionally, the findings presented in Table \ref{tab:DS} shed light on an unexpected outcome when employing low-capacity decoders. With the first set of downstream architectures, the ``Large" versions of SSL models consistently outperform their ``Base" counterparts. However, this pattern does not hold true with higher-capacity decoders in the second set of probes. For example, the best performances in ASV and ER are achieved using WavLM Base+ and HuBERT Base, respectively. In the context of intent classification, changing the downstream decoder from linear to BiLSTMs results in a significant reduction in the mean absolute difference between the ``Base" and ``Large" versions' performance, decreasing from $14.23$ to $3.28$. Again, for emotion recognition, although all four ``Large" versions outperform their ``Base" counterparts with linear probing, increasing the capacity of the probing head reverses this order for all models except WavLM. Additionally, in the case of ASV, DistilHuBERT achieves better results with an ECAPA decoder than the best-performing model (WavLM Large) with an x-vector-based head, despite having more than $13$ times fewer parameters. These findings suggest that using excessively small-capacity heads advantage larger SSL encoders and may have been leading to inflated model sizes.

\subsection{Performance and Inference Costs}
 \begin{figure}[t!]
   \centering
  \includegraphics[width=1\linewidth, scale=0.17]{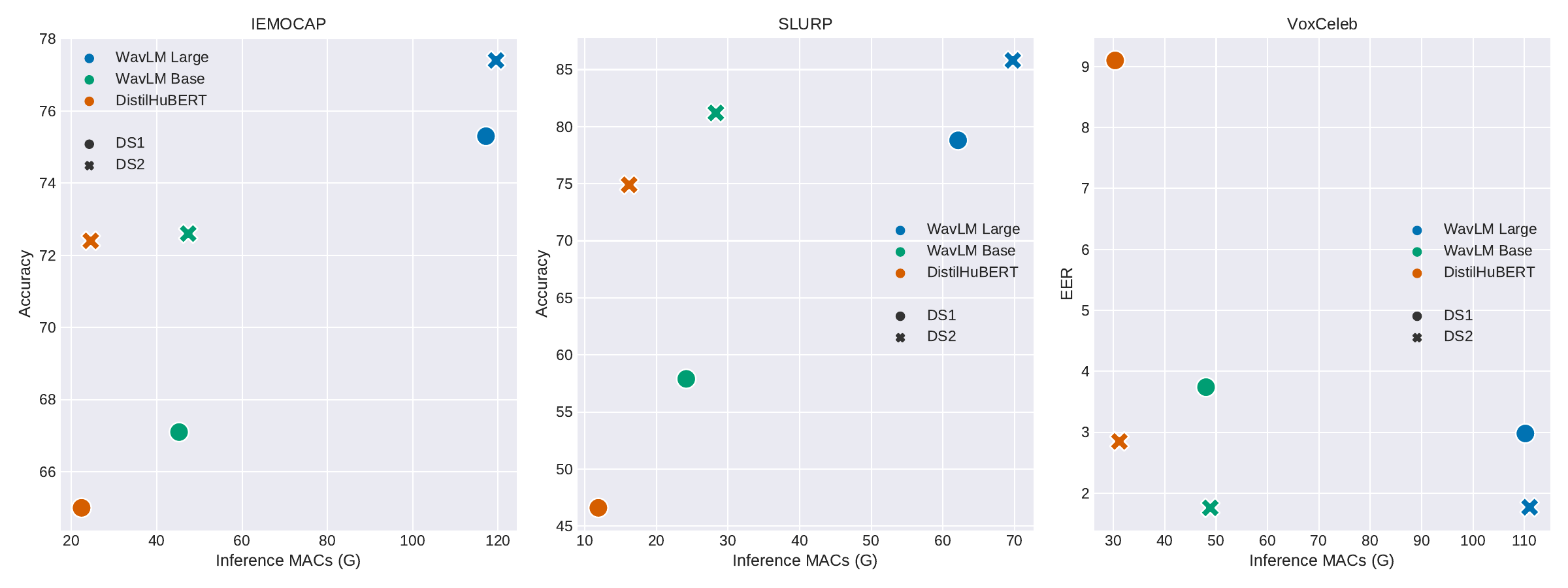}
  \caption{Performance vs mean total inference cost metrics (in G-MACs) depending on the probing heads used for three models and three different downstream tasks. On all tasks, second downstream probes, larger in capacity, allow smaller SSL models to bridge the gap with bigger ones in term of accuracy with limited additional inference costs. $DS(i)$ for $i \in {1,2}$ corresponds to the results obtained with the $i-th$ set of downstream probes. }
   \label{macs}

\end{figure}
 Since the number of parameters does not present a full picture of the computations involved, the THOP library\footnote{\url{github.com/Lyken17/pytorch-OpCounter}} is used to compute the number of Multiply–Accumulate operations (MACs) implied by the learned models. We compute exactly the mean number of MACs involved in inference (self-supervised feature extraction and downstream decoding) for every sample in the test set. Figure \ref{macs} shows the number of inference MACs for three models of different sizes and three considered downstream tasks: emotion recognition, intent classification, and speaker verification. For a fair comparison, we select the large models that perform the best on the considered task with the first downstream probe, along with its ``Base" counterpart and DistilHuBERT as an even smaller competitor. First, on all three tasks, and for every model, the reached performance is systematically better with bigger decoders. Furthermore, the smallest encoder "DistilHuBERT", while bearing 13 times less parameters than ``Large" encoders, reaches a performance with the second decoder that is comparable to the best ``Large" model with the first smaller downstream probe. Visually, for every considered model, the x-axis translation between the ``DS1" (circle-shaped) and ``DS2" points (cross-shaped) shows the MACs quantity increase induced by a bigger decoder head. While the BiLSTM-based decoder is visible on SLURP, the ECAPA-TDNN-based one seems negligible in the two other tasks compared to the self-supervision-based feature extraction costs. The three figures depict clearly both the high performance impact of a small boost in the decoder capacity and its low impact on the total computations needed for inference because of the large cost of feature extraction.

\subsection{Multi-level feature exploitation} \label{sec:layers}
The layer-wise content of speech self-supervised representations has been extensively probed throughout the literature \cite{riera2023phone, pasad2021layer}. These studies generally assess the content with linear probes or with Canonical Correlation Analysis \cite{ankita}.  This subsection studies the impact of changing the probing head on the learned weighting of the layers of the models. It concludes that larger probing heads lead to a better exploitation of multi-level features in the considered self-supervised encoders. 

As stated in section \ref{probes}, during fine-tuning, and in order to cover all the considered downstream tasks, a weighting of the SSL models' layers is learned jointly to the probing heads parameters. With $N$ the number of layers, 1 for the output of the convolutional front-end and $N-1$ transformer layers in the SSL encoders (3 in total for DistilHuBERT, 13 for ``Base" models and 25 for ``Large" ones), $(P_i)_{i \in \{1,.., N\}}$ is a learned vector and $W = Softmax(P)$ is the layer weighting vector. Let $(R_i)_{i \in \{1,.., N\}}$ represent, for a given SSL encoder,  the $N$ matrices of intermediate embeddings of shape $[T,D]$ with $T$ the number of time frames ($50$ per second), and $D$ the dimension of the encoder learned representations. Then the input representation decoded by the probing head is:
\begin{equation}
    R_{input} = \sum_{i=1}^{N} W_i R_i. \\
\end{equation}

 \begin{figure*}[t!]
  \centering
\includegraphics[width=1\linewidth, scale=0.17]{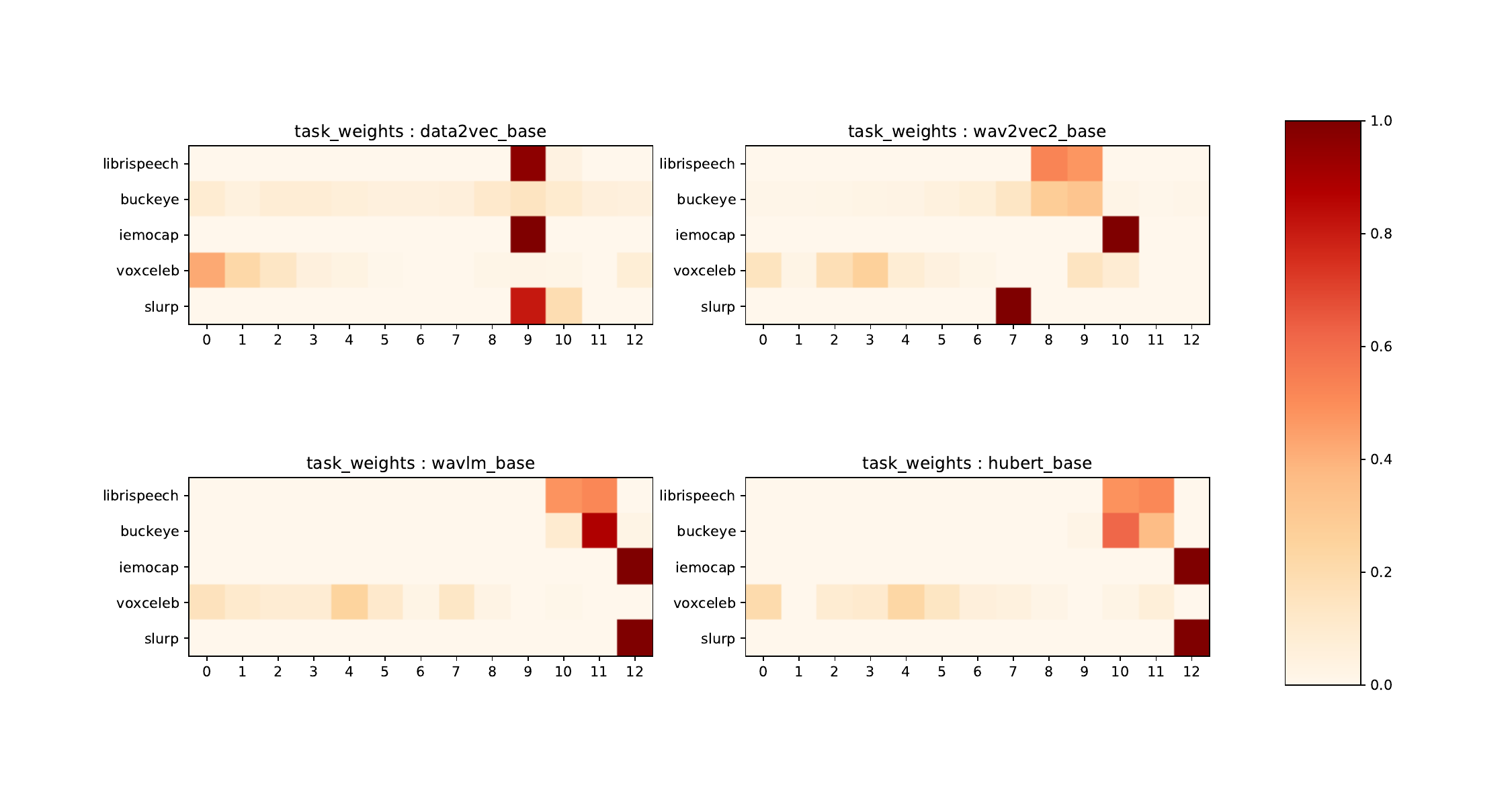}

  \includegraphics[width=1\linewidth, scale=0.17]{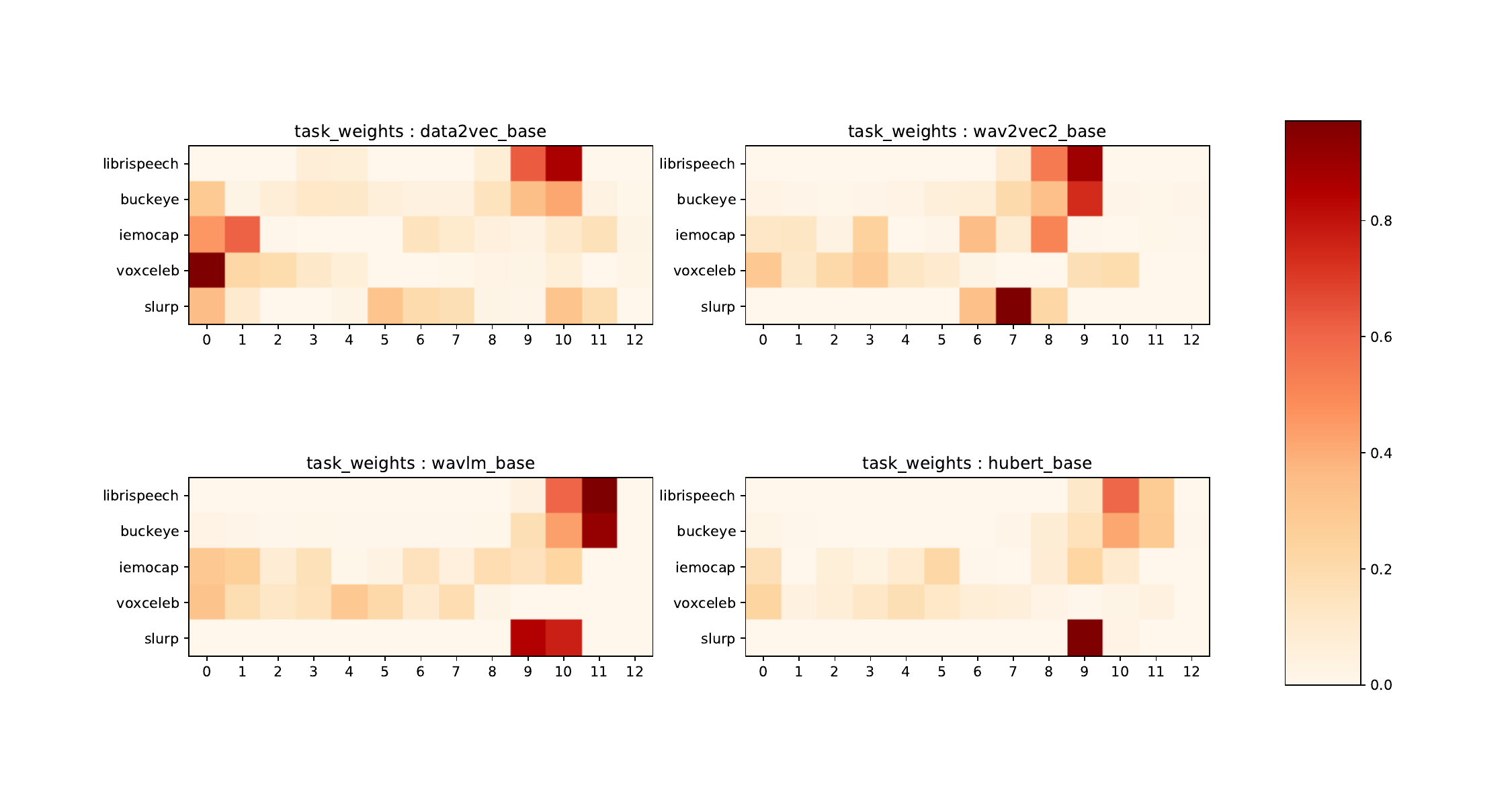}
    \vspace{-1cm}
  \caption{Values of the layer weights learned during fine-tuning for all ``Base" models on the considered tasks. The values on every row sum to $1$. The weights obtained with the second downstream probes (bottom part of the figure) are shifted to lower-level layers compared to the first probes ones (top part).}
  \label{fig:layers2}
\end{figure*}

Figure \ref{fig:layers2} depicts the values (learned during every downstream training) of these weights for the four ``Base" models considered in this work. The top part shows the learned weights with the first downstream probing heads, and the bottom part shows the second ones. First, it is very interesting to observe that the values of the learned weights seem to depend heavily on the SSL encoder pretraining task. While Data2Vec and Wav2Vec2.0 based, respectively, on masked language modeling and contrastive learning of quantized representations, display different weighting, HuBERT and WavLM, that have similar pretraining tasks, have very similar learned weighting for all the considered tasks, and with the two sets of downstream probes. 

Second, it is important to note that the values of the learned weights are heavily impacted by changes in the considered probing head. This is especially the case for non-ASR tasks, and specifically for emotion recognition and intent classification. For these two tasks, with all the self-supervised encoders, only layers above the 9-th are selected with the linear probing approach. However, larger-capacity probes seem to be able to exploit low-level features. 

For IEMOCAP, when using the first probing head, \textit{i.e.} time-pooling followed by a linear classifier, the model relies on features from only one high-level layer (the last one for instance, for HuBERT and WavLM). On the contrary, probing with the ECAPA-TDNN---the second probing head considered here---spreads the weights across the different layers. In some cases, the last layers are barely weighted: Data2Vec, for instance, mainly uses the two first ones as shown in the first plot of the third row in Figure \ref{fig:layers2}. This tends to indicate that the emotion recognition systems built using the linear probe may be exploiting linguistic content, while the second probe exploits mainly low-level emotion-related features.  Concerning intent classification with the SLURP dataset, for HuBERT and WavLM, the main weight moves from the last layer to around the ninth one, while for Data2Vec, the LSTM-based decoder starts using multi-level features, including the first layer, \textit{i.e.} the output of the convolutional front-end. We cannot easily draw a similar conclusion for ASR, where the high-level features are generally the closest to the phonetical content \cite{riera2023phone} and thus to the nature of the ASR task and seem to be naturally preferred by both the considered decoders. Finally, the VoxCeleb speaker recognition is always selecting low-level features, this is coherent with the layer-wise probing literature \cite{pasad2021layer}, showing the loss of speaker information in high-level features of ASR-oriented self-supervised models. 

Building on these observations, we argue that larger-capacity decoders enable the exploitation of multi-level features. In the case of intent classification and emotion recognition, this seems natural  given that the first probes, time-pooling followed by a linear classifier, could only exploit features allowing for linearly separable downstream classes. This multi-level extraction may be behind the substantial increase in performance for both intent classification and emotion recognition. 

We test this conjecture for emotion recognition with another experiment where one downstream probe is learned using fixed weights obtained with the other one. These results are reported in Table \ref{tab:fixedweights}. Precisely, in this experiment, we fix the weights during the downstream training, with the ones obtained either during the first or second probing. In our set of experiments, for every SSL encoder $\phi$, we  learn the parameters of a downstream probing head DS and a set of weights for the layers representations W. In Table \ref{tab:fixedweights}, for every SSL encoder $\phi$, every column $DS(i)/ W(j)$ with $i,j \in {1,2}$, shows the accuracy after decoding with probing head $DS(i)$ but with fixed weights $W(j)$ corresponding to the ones learned initially with $DS(j)$. The results show that, while the larger capacity probing head still performs better than the low capacity ones with their considered weightings, a reasonable part of the performance increase is imputable to the change in the level of features used. With the same ECAPA-TDNN decoder, using multi-level features improves the performance from 68.6 to 73.3 mean accuracy on the 4 SSL encoders considered in this experiment. Another interesting observation is that the first downstream head, time-pooling followed with a linear decoder, is not able to better exploit multi-level features, with very similar performances between the two weightings.
\begin{table}[]
\centering

\scalebox{0.95}{\begin{tabular}{lcccc}
\toprule
\textbf{SSL Model / Head/ Weights} & \textbf{DS1/W1}              & \textbf{DS1/W2}                & \textbf{DS2/W1} & \textbf{DS2/W2}                \\
\midrule
Data2Vec Base & 63 & 63 & 62.6    & \textbf{72.1} \\
Data2Vec Large  & 64                  & 63.9                   & 67.9    & \textbf{71.3}                   \\
WavLM Base    & 67.8                  & 67.9                   & 71.6    & \textbf{72.5}                   \\
WavLM Large   & 75.3                   & 75.3                   & 72.2    & \textbf{77.6}    \\          \bottomrule                  
\end{tabular}}

\caption{Results of experiments on emotion recognition with fixed layer-weights. The result in column $DS(i)/W(j)$ is the one obtained learning the downstream head of the $i-th$ set with fixed weights corresponding to the ones learned originally with the $j-th$ probing head. The difference between column 3 and 4 shows that the exploitation of multi-level features  plays a role in the better performance of DS2.}

\label{tab:fixedweights}
\end{table}

We conclude that probing with larger capacity decoders should be preferred if there is a need to exploit multi-level features, as this  allows for increased performance. We will show in the next section that it also has an impact on generalization on out-of-domain samples. 

\subsection{Generalization Abilities}

A major argument for using low-capacity decoders is that they may allow for better generalization. Indeed, the pre-trained representations are learned on massive amounts of data, with a potential higher data heterogeneity, while the decoding head is learned on small annotated datasets with an expected overfitting hazard. Furthermore, multiple studies have examined and shown the generalization robustness of self-supervised representations \cite{superb_slt,}, which emphasizes, even more, the need to keep this asset. This section aims to show that the models learned with larger capacity decoders are able to generalize as well and even better than their smaller-decoders counterparts, by showing that the performance gains obtained with larger decoders transfer to Out-Of-Domain (OOD) testing samples. Within this scope, we consider the final models obtained with different capacity decoding heads on the considered tasks and test their accuracies on OOD samples, coming from other datasets but having similar downstream classes. This actually enables direct zero-shot generalization performance assessment. Two reasons make two tasks, emotion recognition and speaker verification, relevant for these experiments. First, for both these two tasks, a larger-capacity probing head leads to significantly lower error rates, and we want to test how much this gain is resilient to OOD testing. Second, zero-shot testing requires OOD samples sharing the same labels as the training in-domain set.  For ER, several other datasets share, at least partly, the same labels as IEMOCAP \cite{cao2014crema}. While speaker verification models trained with VoxCeleb \cite{Nagrani_2017} output a binary label indicating whether two samples come from the same speaker or not, and thus can be tested on any other ASV benchmark, including OOD non-English utterances.

 \begin{figure*}[t!]
  \centering
  \includegraphics[width=1\linewidth, scale=0.17]{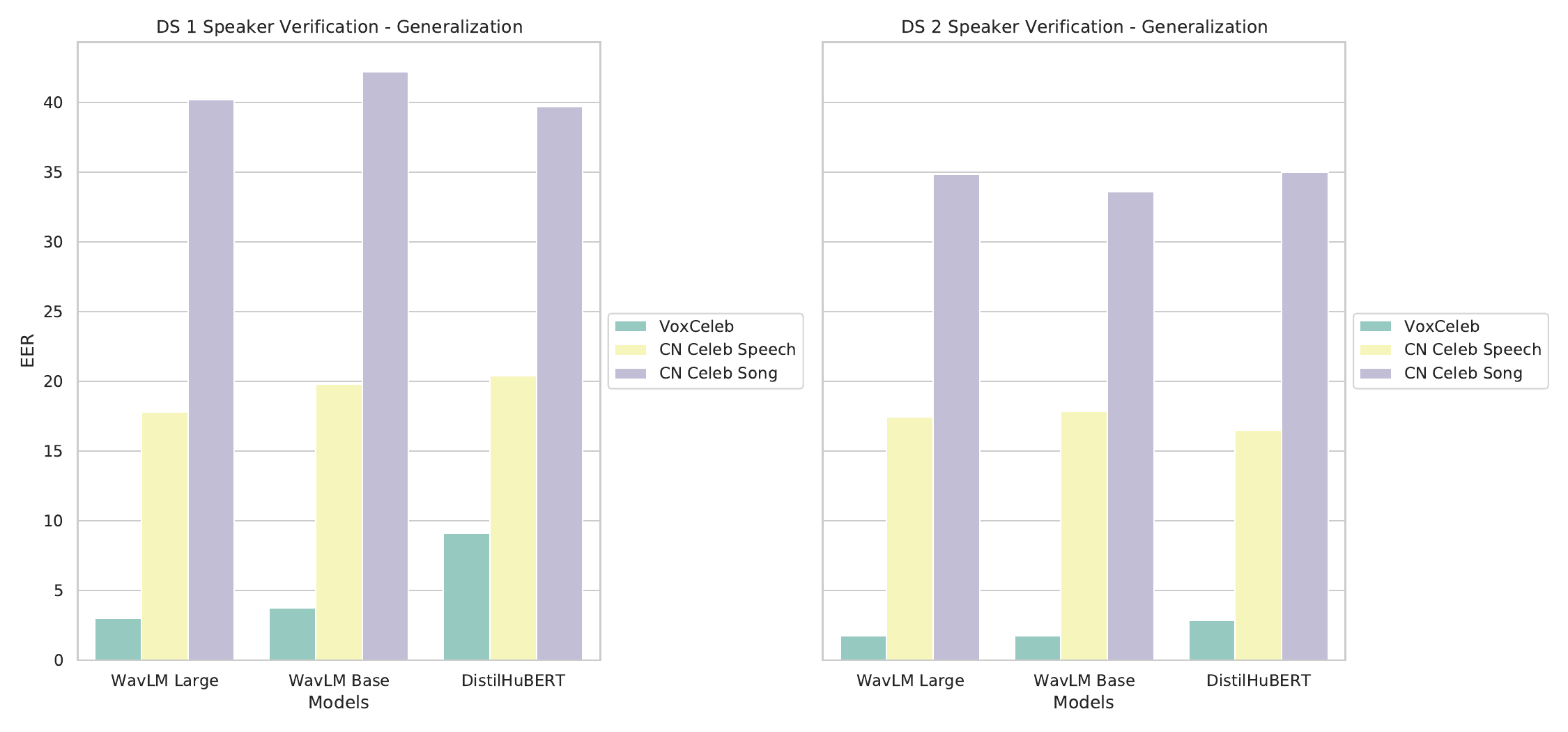}
  \caption{Generalization performances for automatic speaker verification. CN-Celeb Speech and CN-Celeb Song performances are provided in a zero-shot generalization setting and are not included in the training set. Random performance is at 50 EER, and is not shown for better visualization. Larger probing heads, here ECAPA-TDNN, shown in the right plot, generalize better to out-of-distribution testing samples.}
\label{fig:asv_generalization}

  \end{figure*}

 \begin{figure*}[t!]
  \centering
  \includegraphics[width=1\linewidth, scale=0.17]{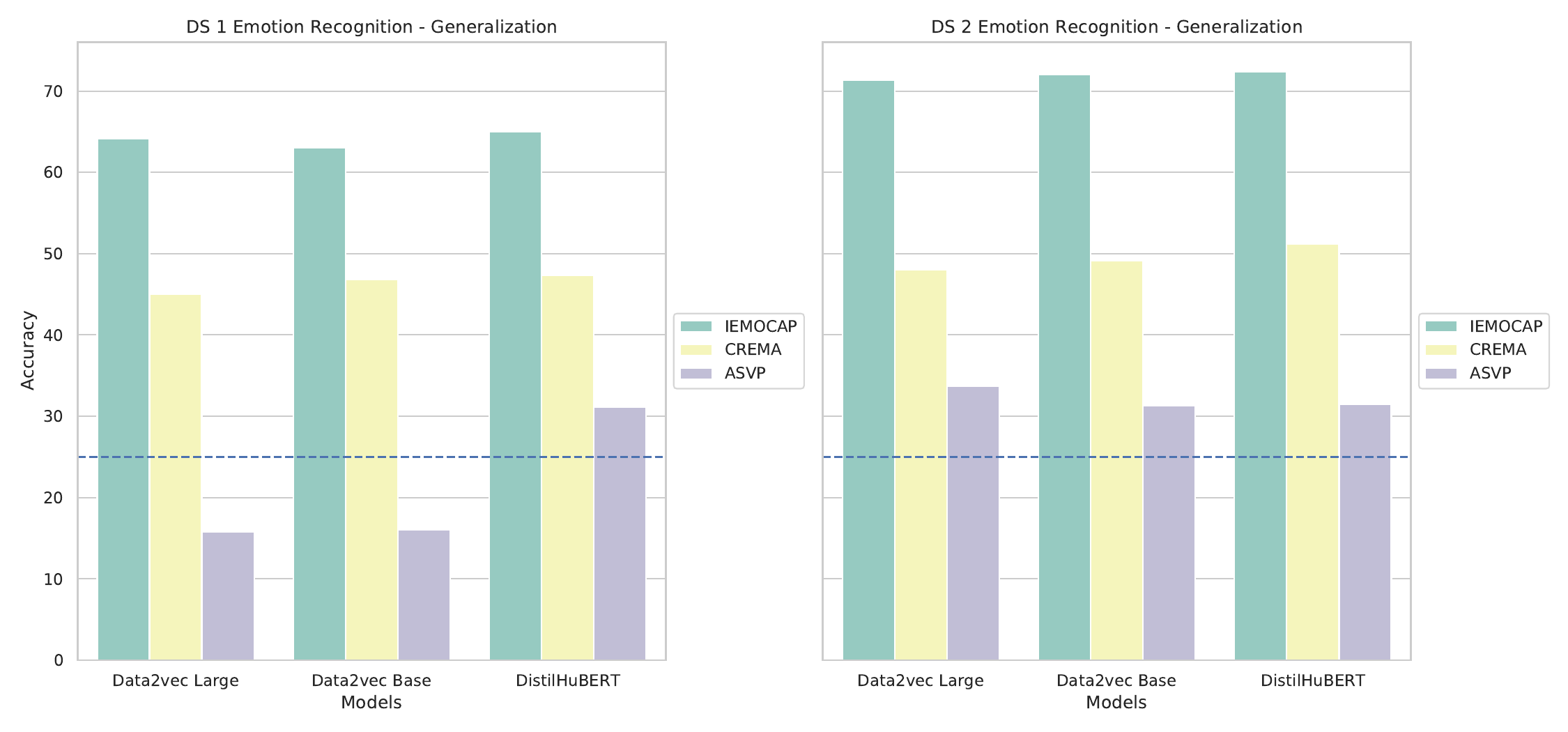}
  \caption{Generalization performances for emotion recognition. CREMA-D and ASVP-ESD performance is tested in a zero-shot setting. The dashed blue line represents the random accuracy level. Larger probing heads, here ECAPA-TDNN, shown in the right plot, generalize better to out-of-distribution testing samples.}
\label{fig:er_generalization}

\end{figure*}

\noindent \textbf{Emotion recognition}. To test the generalization abilities of models learned with different decoders, and after training with IEMOCAP as described in Section \ref{sec:setting}, we test the models in a zero-shot fashion, without further fine-tuning, on two datasets: CREMA-D \cite{cao2014crema} and ASVP-ESD \cite{tientcheu_touko_landry_dejoli_2021_7132783}. CREMA-D is a data set of $7,442$ original clips from $91$ English-speaking actors reading sentences using one of six different emotions (\textit{Anger}, \textit{Disgust}, \textit{Fear}, \textit{Happiness}, \textit{Neutrality}, and \textit{Sadness}). ASVP-ESD is a multi-authentic emotional corpus sourced from movies, Youtube channels, and real-life human interactions in natural settings, without any language limitations. The corpus comprises $5,146$ samples, with $60\%$ consisting of non-speech emotional sounds and $40\%$ comprising speech utterances. For both datasets, only speech elements with labels overlapping with the four IEMOCAP ones (\textit{Angry, \textit{Happy}, \textit{Neutral}, Sad}) are considered. For these two corpora, the testing sets are of reduced sizes. So to increase the significance of the reported results, and since the train sets are not used for training, all the splits (train and test) are used for testing. For ASVP-ESD, and to further enforce OOD testing, English samples are removed. 

\noindent \textbf{Automatic speaker verification}. For speaker verification, the generalization abilities of the models learned on VoxCeleb1, are tested for two out-of-domain scenarios, also in a zero-shot transfer setting. For this, The CN-Celeb dataset \cite{fan2020cn}, a comprehensive collection of speaker recognition data, is used. It encompasses over 130,000 utterances from 1,000 Chinese celebrities, spanning 11 diverse genres (interviews, movies, songs...). To further highlight generalization ability, we divide CN-Celeb testing couples into ones that include one singing voice element, and once with only spoken utterances, leading to two generalization testing sets: ``CN Celeb Speech" and ``CN Celeb Song". The second split is even more challenging in our case, as no singing voice is included in VoxCeleb.

\noindent \textbf{Discussion.} Figures \ref{fig:asv_generalization} and \ref{fig:er_generalization} show the results of these experiments for models built on certain considered SSL encoders. We can note, first, the expected considerable performance loss on the OOD samples, and especially the loss when changing the ER language with ASVP-ESD or testing on singing voice speaker verification with ``CN Celeb Song". For both tasks, as stated in previous sections, in domain performance, \textit{.i.e} performance on the test sets of the downstream training datasets, obtained with the second set of larger probing heads are higher than those with SUPERB limited-capacity probes. The two figures further show that this performance gap stands for zero-shot generalization. Concerning emotion recognition, the mean accuracy on the three considered models reaches $49.43$ and $32.17$ respectively on CREMA-D and ASVP-SED with the ECAPA-TDNN probing head compared with $46.37$ and $20.97$ with the time-pooling followed with a linear decoder.  For speaker verification, enhancing the probing head drives the Equal Error Rate on the ``CN Celeb Speech"  from $19.34$ to $17.27$, while it goes from $40.68$ to $34.46$ on ``CN Celeb Song". In subsection \ref{sec:layers}, we hypothesized that ER models with the first downstream probes may be using linguistic information since only high-level layers were used. The big drop in performance on ASVP-ESD of Data2Vec ``Base" and ``Large" models goes in that direction. Changing the language of the inputs leads to catastrophic performance drops. This is not the case for DistilHuBERT as the model only contains three layers. These experiments show that the gain in performance is not only relevant to in-domain data, but models built on top of frozen SSL encoders reach better out-of-domain zero-shot accuracies with larger-capacity probing heads.

\section{Headless Alternatives}\label{sec:headless}
We have shown in the previous sections that the current methods for benchmarking self-supervised representations were dependent on the downstream head choice. This section explores benchmarking approaches that get rid of this component, as it would solve the problem raised with this dependency. Thus, we discuss approaches to evaluate speech representations without training downstream heads, in what we call here ``headless" evaluation of speech self-supervised models. Precisely, we will define two methods for the headless assessment of the quality of speech representations for two tasks: automatic speech recognition (ASV) and speaker verification (SV). While these methods are headless, in the sense that they do not require a downstream architectural choice or even downstream training, we will see that they still require several other choices and decisions in the definition of the method and its hyperparameters.

\subsection{ABX Tests for Speech Representations}

The ABX test \cite{schatz} assesses the discriminability of speech representations.  In a single ABX test, three audio segments are considered: \(A\), \(B\), and \(X\). $A$ and $X$ are picked so that they share the same downstream identity or label. For instance, if we are probing for speech recognition, $A$ and $X$ should hold the same phonetic content. If the downstream task considered is speaker-related, then two speech segments pronounced by the same speaker should be selected. In contrast, $B$ bears a different downstream label, \textit{i.e.} in the two examples above, a different phonetic sequence, or a different speaker. 

A self-supervised encoder, here a function $\phi$, is evaluated by checking whether $\phi(X)$ is more similar to $\phi(A)$ or $\phi(B)$. Mathematically, the decision is based on a comparison of the similarity scores between $\phi(X)$ and $\phi(A)$ (\textit{i.e.} \(S(\phi(X), \phi(A))\)) and $\phi(X)$ and $\phi(B)$ (\textit{i.e.} \(S(\phi(X), \phi(B))\)), with $S$ a similarity measure. If the first couple is more similar than the second, it means that the encoder maps the audios that share the same downstream label to points closer in the representation space than one not sharing the same label, implying that the self-supervised representations allow for the separation of the classes of interest. Multiple tests, using different $(A,B,X)$ triplets, are produced for every task. The scores that are shown are the error rates for a given set of triplets, \textit{i.e.} the proportion of triplets where $\phi(X)$ is more similar, according to the measure $S$, to $\phi(B)$ than to $\phi(A)$.

\subsection{Speech Recognition}
We mine a set of $10000$ triplets following the instructions presented by Nguyen \textit{et al.} \cite{DBLP:journals/corr/abs-2011-11588} extracted from the Buckeye dataset presented in Section \ref{sec:setting}. Since two segments, even representing the same phonetic sequence, do not necessarily have the same length, we chose, again as in \cite{DBLP:journals/corr/abs-2011-11588}, for the similarity measure  $S$, the average cosine similarity of the representations along a realigned path obtained with Dynamic Time Warping (DTW) \cite{10.5555/3000850.3000887}.  We provide in the following a more detailed description. Let us consider two sequences $A$ and $B$, where $A = [a_1, a_2, ..., a_n]$ is the concatenation of $n$ frames representations and $B = [b_1, b_2, ..., b_m]$ similarly the concatenation of $m$ ones. We also consider the  normalized dot product (or cosine) as the similarity measure between two representations: 
$$s(a_i, b_j) =  \frac{a_i  \: .  \: b_j}{|a_i| \: . \: |b_j|} \ . $$
The sequence level similarity measure is then: 

$$S(A,B)= D T W(A, B)=\min _{\pi \in \mathcal{A}\left(A, B\right)}\left(\sum_{(i, j) \in \pi} s\left(A_i, B_j\right)\right)$$ 

where $\mathcal{A}(A,B)$ is the set of all admissible warping paths between the sequences $A$ and $B$. We will be comparing the rankings and relative performances obtained with this probing method to the ASR values in Table \ref{tab:DS}.  

\subsection{Speaker Verification}
The speaker verification task is already close to an ABX task. The difference is that it is rather an AX one \cite{kamper2016deep}, with only two segments considered, and the use of a similarity threshold for decision, instead of a comparison. In detail, with $\phi$ again the self-supervised encoder and $f$ the downstream learned speaker head, given two audio segments $A$ and $X$, a similarity measure $S(f(\phi(A)), f(\phi(X)))$ is computed, and a positive match is predicted if $s$ is above a certain threshold $\tau$. Nonetheless, in self-supervised speaker verification settings, no downstream speaker head is learned, and the self-supervised representation is used directly in the similarity computation becoming $S(\phi(A), \phi(X))$. The threshold $\tau$ is selected as to minimize the Equal Error Rate (EER). 

We have seen in Section \ref{sec:layers} that the speaker downstream heads tend to use low-level features in the self-supervised encoder. Taking this into account, we do not use the final output of the encoder, but rather a weighted mean with exponential decay with respect to the depth of the layer, \textit{i.e.} allowing more weight to early layers. The weights are computed according to the following formula: $W = softmax( (exp(- \lambda *i))_{i \in [1, N]})$ with $N$ the number of layers in the SSL encoder and $\lambda$ a decaying hyperparameter fixed to $\lambda=0.2$. Again, as with speech recognition, we face the problem of varying length sequences. To avoid these, we use the same similarity $S$ described above. It allows us to compute an equal error rate (EER) similarly to what has been done in Section \ref{sec:setting}, but without a speaker-trained downstream head. We use VoxCeleb1 enrolment and test trial couples, as in the previous sections.
\begin{table}[] 
\centering
\begin{tabular}{lcccc}\toprule
Model             & \multicolumn{2}{c}{\textbf{Buckeye}}        &   \multicolumn{2}{c}{\textbf{VoxCeleb1}}          \\ 
                  & ASR (WER)  & ABX (ER)   & SV (EER)  & AX (EER) \\ \midrule
DistilHuBERT      & 58.56 (9) & 15.68 (9)& 2.85 (6)& 15.12 (9)\\
Wav2vec 2.0 Base  & 30.04 (5)& 13.48 (8)& 2.82 (5)& 13.34 (6)\\
Wav2vec 2.0 Large & 23.92 (2) & 11.57 (7)& 3.17 (7)& 9.30 (3)\\
HuBERT Base       & 30.44 (6) & 10.90 (5)& 2.40 (3)& 13.42 (7)\\
HuBERT Large      & 39.39 (7) & \textbf{8.44 (1)} & 3.84 (9)& \textbf{8.70 (1)} \\
WavLM Base+       & 27.73 (4)& 10.75 (4) & \textbf{1.76 (1)} & 12.87 (5)\\
WavLM Large       &\textbf{ 15.61 (1)} & 8.61 (2)  & 1.77 (2) & 8.73 (2)\\
Data2Vec Base     & 40.53 (8)& 11.43 (6)& 3.75 (8)& 14.10 (8)\\
Data2Vec Large    & 25.56 (3)& 9.17 (3) & 2.67 (4)& 10.07 (4)\\ \bottomrule
\end{tabular}
\caption{Results obtained when evaluating with and without downstream heads for two considered tasks, speech recognition on the Buckeye dataset and speaker verification with VoxCeleb1. The ``ABX" and ``AX" columns indicate the ``headless" evaluations. We witness considerable differences in the rankings and relative performances between the two types of evaluations. For every metric, lower values indicate better performance. The ranking of the model in each column is shown between parenthesis.}
\label{tab:headless}
\end{table}
\subsection{Results and Analysis}
Table \ref{tab:headless} shows the results obtained with the described approaches. We copy the results obtained for ASR and SV, obtained with the second set of downstream heads in Section \ref{sec:results}. We compute again the Pearson and Spearman correlation for the values obtained with two different evaluation techniques for every task. The correlations between the ASR and ABX results reach, respectively, $0.67$ and $0.48$ on the Buckeye dataset. For the speaker task, the correlations between the SV Equal Error rate and the AX task error rate, attain -0.01 for Pearson and -0.02 for Spearman. The correlation values for the Buckeye dataset resemble those obtained in the previous sections. They are positive, showing that the evaluation metrics are linked, but not enough to validate the robustness of the evaluation method. For VoxCeleb1, the very low values show that the two evaluation methods are not correlated, suggesting dropping the use of the AX approaches for the evaluation of speaker content in speech self-supervised representations. 

While headless approaches are appealing, as they cancel the dependency on the architectural choice for the downstream heads, the correlation values show that they do not correlate highly with the performance on the exact downstream tasks with complex probes. The results are not surprising as the ABX or AX evaluation methods can be seen as some sort of low-capacity probing of the self-supervised representations. Indeed, a good performance on these headless tasks indicates that the downstream labels of interest, either phonetic or speaker identities in our cases, are directly clustered in the embedding space. If they are, then their separation should be accessible without complex further feature processing. 

Finally, we should highlight that these techniques also involve hyperparameters and choices, essentially concerning the triplet mining and the similarity function used. Differences in the choice of the similarity function may lead to very different rankings, leading to the same problem observed for downstream architectures.

\section{Acknowledgements}
This work has benefited from funding from l'Agence de l'Innovation de Défense, and was performed using HPC resources from GENCI-IDRIS (Grant 2021-AD011012801R1).

\section{Conclusion}
It is crucial to improve the way the speech community currently benchmarks widely used self-supervised representations, first, because better benchmarks allow SSL users to select properly the models they need for their downstream tasks of interest. Second, it offers the SSL model developers insightful evaluations shaping the training process and decisions. In this work, we have shown, by varying the downstream architectures, that the ranking and relative performances of popular self-supervised models heavily depend on the choice of the probing heads. While the community has previously chosen to evaluate the learned representations with limited-capacity decoders, we have revealed, as an additional contribution, that larger-capacity decoders should be preferred in various scenarios. This is motivated by better performances, a reduced performance gap between  ``Base" and ``Large" encoders leading to high (performance/inference costs) ratios, better multi-level feature exploitation, and better out-of-distribution generalization. We hope this diagnosis will support the community in designing new benchmarking approaches and encourage submissions to the SUPERB ``Constrained" track described in the introduction or propose new probing heads in the dedicated benchmark section within the SpeechBrain Library.

\bibliographystyle{IEEEtran}
\bibliography{mybib}

\vfill

\end{document}